\documentclass{mem}
\usepackage{natbib}\usepackage{txfonts}\usepackage{balance}
\usepackage{graphicx}
\usepackage{txfonts}
\begin{document}
\newcommand{\scrbox}[1]{\ensuremath{{\mbox{\scriptsize #1}}}}

\newcommand{\teff}{{\ensuremath{T_{\scrbox{eff}}}}}
\newcommand{\Msol}{\ensuremath{\,\mbox{\it M}_{\odot}}}
\newcommand{\Mstar}{\ensuremath{\it \,M_{*}}}
\newcommand{\Mmixed}{\ensuremath{\Delta M_{\rm m}}}
\newcommand{\Dturb}{\ensuremath{D_{\scrbox{T}}}}
\newcommand{\Kelvin}{\,\mbox{K}}
\newcommand{\MS}{main--sequence}
\newcommand{\gr}{\ensuremath{g_{\scrbox{rad}}}}
\newcommand{\tbcz}{{\ensuremath{T_{\scrbox{bcz}}}}}
\newcommand{\mbcz}{{\ensuremath{M_{\scrbox{bcz}}}}}
\newcommand{\DM}{\ensuremath{ \log \Delta M/M_{*}}}
\newcommand{\DMsol}{\ensuremath{ \log \Delta M/M_{\odot}}}

\renewcommand{\H}{\mbox{H}}
\newcommand{\He}{\mbox{He}}
\newcommand{\Be}{\mbox{Be}}
\newcommand{\B}{\mbox{B}}
\newcommand{\Fe}{\mbox{Fe}}
\newcommand{\Mn}{\mbox{Mn}}
\newcommand{\Mg}{\mbox{Mg}}
\newcommand{\Si}{\mbox{Si}}
\newcommand{\Cr}{\mbox{Cr}}
\newcommand{\Ca}{\mbox{Ca}}
\newcommand{\Ni}{\mbox{Ni}}
\newcommand{\Ne}{\mbox{Ne}}
\newcommand{\Ti}{\mbox{Ti}}
\newcommand{\F}{\mbox{F}}
\newcommand{\Ox}{\mbox{O}}
\newcommand{\K}{\mbox{K}}
\newcommand{\Li}{\mbox{Li}}

\newcommand{\mix}{_{\rm{mix}}}
\newcommand{\T}{_{\rm{T}}}
\newcommand{\ad}{_{\rm{ad}}}
\newcommand{\rad}{_{\rm{rad}}}
\newcommand{\Le}{_{\rm{L}}}
\newcommand{\SC}{_{\rm{SC}}}
\title{
Radiative Accelerations in Stellar Evolution
}

   \subtitle{}

\author{
G. \,Michaud\inst{1,2} 
\and J. \,Richer.\inst{1}
          }

  \offprints{G. Michaud}

\institute{
D\'epartement de Physique, Universit\'e de Montr\'eal,
       Montr\'eal, PQ, H3C~3J7
\and
LUTH, Observatoire de Paris, CNRS, Universit\'e Paris Diderot,
     5 Place Jules Janssen,
     92190 Meudon, FRANCE
     \email{michaudg@astro.umontreal.ca; jacques.richer@umontreal.ca}
}

\authorrunning{Michaud}

\titlerunning{Radiative accelerations}

\abstract{
A brief review of various methods to calculate radiative accelerations for stellar evolution and an analysis of their limitations are followed by applications to Pop I and Pop II stars.  Recent applications to Horizontal Branch (HB) star evolution are also described.  It is shown that models including atomic diffusion satisfy Schwarzschild's criterion on the interior side of the core boundary on the HB without the introduction of overshooting.  Using stellar evolution models starting on the Main Sequence and calculated throughout evolution with atomic diffusion, radiative accelerations are shown to lead to abundance anomalies similar to those observed on the HB of M15. 
}
\date{}
\maketitle{}

\section{Introduction}
In his book, \citet{Eddington26} evaluated the equilibrium concentrations to be expected if atomic diffusion in the presence 
of differential radiation  pressure were efficient.  He concluded that some heavy metals should then completely dominate the spectrum. 
 Since light and heavy elements are present in stellar spectra, he concluded (\S 193, 196) that some mixing process made atomic diffusion inefficient.
  He suggested that it be meridional circulation (\S 199).  
His argument for the importance of meridional circulation was qualitative.  In a later paper he evaluated
 quantitatively the meridional circulation velocity without further commenting 
on its efficiency in mixing stellar interiors \citep{Eddington29}.  While it was realized in the 1940s that the presence of the giant branch 
in clusters implied that stars could not be completely mixed, Eddington's argument seems to have had enough weight to prevent the proper 
calculation of the effect of 
differential radiation pressure until the late 1960s even if Eddington had partially corrected his argument (see the 1930 correction on page xiii of \citealt{Eddington26}). Work however continued on gravitational 
settling in outer solar layers \citep{Biermann37,Wasiutynski58,AllerCh60}; the most important application was made to white dwarfs \citep{Schatzman45}.  

Instead of assuming that equilibrium concentrations were reached, one of us introduced the differential radiative acceleration term, \gr, into
the diffusion velocity equation of \citet{AllerCh60} and calculated anomalies to be expected in atmospheric regions \citep{Michaud70}.  Comparison to observations of ApBp stars suggested that \gr{} plays a role in at least some stars.
As large atomic data bases started becoming available in the 1980s, we realized that it would be possible to calculate
\gr{} throughout stellar interiors with good accuracy at the same time as the evolution proceeds.  All composition changes can then be taken into account self consistently during evolution for the \gr{} and  the Rosseland
averaged opacity.  It was first tried to make those calculations with TOPBase from the Opacity Project \citep{AlecianMiTu93,LeBlancMi95,GonzalezArMi95,GonzalezLeAretal95} but using those data we could not reproduce the concentration dependence of the Rosseland averaged OPAL opacities around the solar center (see \S 5.1 of 
\citealt{TurcotteRiMietal98}) and we shifted to using OPAL spectra \citep{RicherMiRoetal98} to calculate \gr{}.  The original spectra they had used to calculate the OPAL opacity tables   \citep{IglesiasRo93,IglesiasRo96,RogersIg92a} 
were kindly made available to us by Iglesias and Rogers.

In this  review of \gr{} in stellar evolution, we will first briefly describe how the calculations are carried out in 
stellar evolution codes (\S\,\ref{sec:calculations}) and compare the values of \gr{} obtained using OP and OPAL data (\S\,\ref{sec:CorrectionFactors}).  A few examples of the effect 
of \gr{} in Pop I and Pop II stars are described (\S\,\ref{sec:Role}) and, 
  finally,  results obtained recently for HB stars are presented (\S\,\ref{sec:HBStars}) showing that \gr{}s play a role in stellar evolution (\S\,\ref{sec:con}).

\begin{figure*}[t!]
\resizebox{\hsize}{!}{\includegraphics[clip=true]{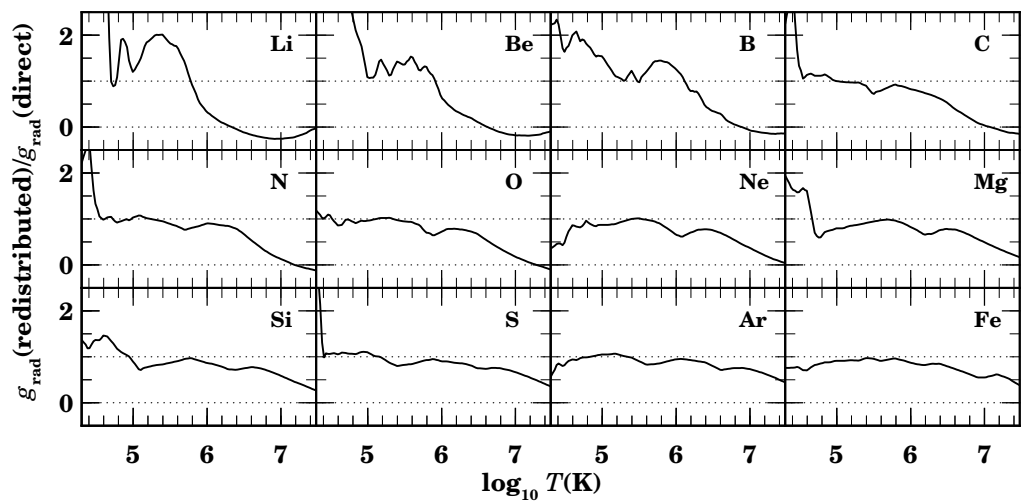}}
\caption{\footnotesize
Correction factors used in evolution calculations except for Fe, see the text. Adapted  from Fig. 6 of \citealt{RicherMiRoetal98}.
}.
\label{fig:correction}
\end{figure*}

\section{Calculations of radiative accelerations}
\label{sec:calculations}
The particle transport equations are introduced into a standard stellar evolution code descibed in 
\citet{Proffitt94}, \citet{ProffittMi91} and \citet{VandenBerg85}.
For each species one adds a force equation (eq. [18.1] of \citealt{Burgers69}) and a heat equation (eq. [18.2] of Burgers). Similar equations are written for electrons. 
 It is generally assumed that each atomic species can be treated  locally as being in an average state of ionization.
One  needs to know $Z_i$, an appropriate mean of the number of lost electrons.

The dominant term for each species contains $\gr(A) - g$ as a factor, where $\gr(A)$ is an appropriate average 
of the radiative acceleration over the states of ionization of element $A$.  Over most of the stellar interior and for most of the evolution, it dominates transport even though the electric field, ``diffusion'' and thermal diffusion terms are also included in the calculations and are important for part of the evolution in some stars. 

Rosseland opacities, mean ionic charges, and mean radiative forces are calculated
using the same interpolation method, based on the principle of corresponding
states in  \S\,2.2 of \citet{RogersIg92a}; see in particular their equations (4) to (6).   Interpolation
weights are determined for a subset of the data grid, and used to interpolate locally all
these variables.

In first approximation, evaluating  $\gr(A)$ amounts  to
calculating the fraction of the momentum flux that each element absorbs from the photon flux.  In
stellar interiors, it takes the form:
\begin{equation}
\gr(A) = {{L_r^{\mbox{\scriptsize rad}}}\over{4\pi r^2 c}}
   \frac{\kappa_{R}}{X_A} \int_0^\infty
   \frac{\kappa_u(A)}{\kappa_u(\mbox{total})} { \cal{P}}(u)du
 \label{eq:grad_simple_def}
\end{equation}   
where most symbols have their usual meaning.  The quantities  $\kappa_u(total)$ and $\kappa_u(A)$
are respectively the total opacity and the contribution of element $A$ to the total opacity at frequency $u$, with  $u$ and ${\cal{P}}(u)$  given by:
\begin{equation}
u={{h\nu}\over{kT}}
\label{eq:u}
\end{equation}
and
\begin{equation}
{\cal{P}} \left({u}\right)={{15}\over{4\pi^4}}  {u^{4}{{e^{u}}\over{{\left({e^{u}-1}\right)}^{2}}}}.
\label{eq:pu}
\end{equation}
The calculations of \gr(A) involve carrying out the integration over the 10$^4$ $u$ values for each atomic species, A.  This must be repeated at each mesh point (typically 2000 in our models) and at each time step (typically 10$^4$ up
to the HB). The Rosseland average opacity is also continuously recalculated making these calculations fully self consistent with all composition changes. This has some effect on stellar cores (see Fig. [9] and the last two paragraphs  of \S 5.1 of \citealt{TurcotteRiMietal98} for a discussion of the solar case).

\subsection{Correction factors}
\label{sec:CorrectionFactors}
The \gr{}s of equation (\ref{eq:grad_simple_def}) are corrected by two processes.  First by taking into account the 
fraction of momentum that is given to the electron in a photoionization, as first suggested in this context 
by \citet{Michaud70} following \citet{Sommerfeld39}, and second by calculating the effect of having many stages
of ionization and not only an average one, which has come to be called the effect of redistribution.

The sharing of the momentum between the ion and the ejected electron is caused by the process remembering the direction of the incoming photon and emitting the electron with a distribution which is not exactly spherical.  The correction to sphericity is small ($\sim \upsilon/c$ where $\upsilon$ is the electron velocity) but for a given energy, the momentum of the electron 
being $c/\upsilon$ times that of the photon, the correction to momentum transfer is large.  The effect is difficult to calculate except for the ground state of  hydrogen  but accurate evaluations now exist for a few other cases and 
confirm the generalization formulas which were used \citep{Massacrier96,MassacrierEl96,Elmurr99,Seaton95}.
Using their results, \citet{RicherMiMa97} showed that for any shell $n$ of an hydrogenic ion, one could use the simple formula for the effect of momentum sharing \citep{Michaud70}:
\begin{equation}
f_{\mathrm{ion}}(n) = 1 - \frac{8}{5}\left(1-\frac{{\nu}_n}{\nu}\right),
	\label{eq:fion}
\end{equation}
which has been used for all cases calculated up to now.

When many states of ionization are present, one usually works, for convenience, with an average state of ionization.
The simplest ``averaging'' is to use a single diffusion velocity for each atomic species and calculate it for the average $Z$ of the atomic species, 
 \begin{equation}
	\overline{Z} = \frac{\sum_{i}X(A_i) Z_i}{\sum_i Z_i},
	\label{eq:averagedZ}
\end{equation}
where the sum $i$ is over the states of ionization of element $A$.
The use of equation (\ref{eq:grad_simple_def}) is in line with this approach.  However each term appearing in a diffusion equation has its own $Z$ dependence and this approach is not always satisfactory.  The \gr{} term is the most
 sensitive to the averaging process since very often the largest $\gr(A_i)$ is for a very unabundant ion 
 whose lines are desaturated.
One could use an average of the form:
\begin{equation}
<g_{\mathrm{rad}}(A)>=
\frac{\sum_i D({A_i}) X(A_i) \gr(A_i)}{\sum_i D({A_i}) X(A_i)}
\label{eq:gm}
\end{equation}
where $\gr(A_i)$ would be due to the fraction of the photon flux momentum absorbed while the atomic species is 
in state $i$. However this assumes that one may consider each ionic state of the atomic species $A$ as independent.
In practice this is a poor approximation since there are frequent ionizations and recombinations.  Consequently 
one needs to compare the collision and ionization times and take into account the various ionization routes.  This rapidly becomes complex and has been solved in detail only for He \citep{MichaudMoCoetal79} and Hg \citep{ProffittBrLeetal99} with additional calculations described in \citet{GonzalezArMi95,GonzalezLeAretal95} for CNO.

In the evolution calculations these two effects, momentum sharing
between ion and electron, and averaging over the ions, are combined into a multiplicative factor ($=[\gr(\rm {redistributed})/\gr(\rm {not \, redistributed})])$, which we call a correction factor, applied to the result of equation (\ref{eq:grad_simple_def}).
It was calculated for a number of species (see Fig. [\ref{fig:correction}]) using 
 OP data for solar composition and tabulated as function of $R_e\,(\equiv N_e/T^3)$ and $\log  T$.  They are used for all 
 compositions. Those shown in Figure \ref{fig:correction} were calculated at $\log R_e = 2.5$ which
corresponds approximately to densities within \MS{} models. One notes that the correction factors are close to 1.0
for $5 < \log T< 6$.  This is the $T$ interval over which \gr{}s play the main role in evolution calculations
done up to now.  

For $\log T > 6$ the main contribution to the correction factor comes from the sharing of momentum with the electron. For all species shown except Fe the correction close to the center comes from hydrogenic ions for which it is believed to be reasonnably accurate.  
It was decided not to apply the correction to Fe following the argument of \citet{Seaton97} that $f_{ion}$ should not be applied to autoionization resonances which often dominate 
in non hydrogenic ions and especially for Fe.
For $\log T < 5$ the correction can be large and should be considered uncertain when they exceed a factor of 2.  This usually occurs where the $\gr(A_0)$ contributes a large fraction of the value obtained with equation (\ref{eq:gm}).
Since the diffusion coefficient of the neutral state is much larger than those of the ionized states \citep{MichaudMaRa78}
a small concentration of the neutral can lead to a large effect which however depends on the dominant ionization processes \citep{MichaudMoCoetal79}.

\subsection{Radiative accelerations from OP data}
\label{sec:GrFromOPData}
In so far as we know, all stellar evoluton calculations done  with \gr{} have been done using OPAL data for equation (\ref{eq:grad_simple_def}) and the correction factors described above calculated using OP data.  However  it is 
possible to do all calculations with OP data using the spectra available from their server \citep{Seaton2005,MendosaSeBuetal2007}.
Those spectra use 10$^4$ frequency values just as OPAL spectra but they are equally spaced in a modified frequency which takes the local flux intensity into account (see \S 2.2.1 of \citealt{Seaton2005}).  It is similar to that proposed by 
\citet{LeBlancMiRi2000} and should have a similar effect on accuracy as discussed there. This mesh should lead to greater accuracy for $\log T < 5$.  However OP has two disadvantages.  
First it has fewer atomic species than OPAL  and the 
second point is that their data base does not contain the $\overline{Z}$ which is needed to calculate the diffusion velocities.  These would need to be calculated separately.

\begin{figure}[t!]
\resizebox{\hsize}{!}{\includegraphics[clip=true]{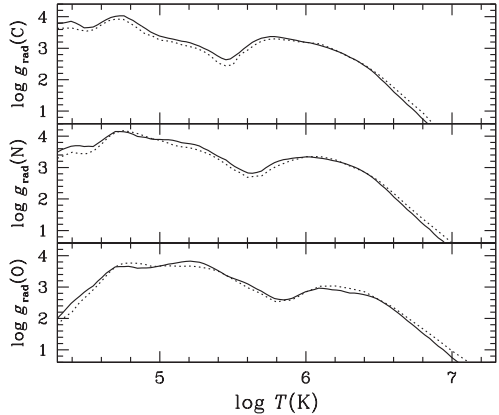}}
\caption{
\footnotesize
Comparison of radiative accelerations calculated from OP (dotted line) and OPAL (solid line) atomic data. From Fig. [7] of \citealt{RicherMiRoetal98}.}
\label{fig:OP_OPAL}
\end{figure}
A comparison of \gr{} calculated with OP atomic data to similar calculations with OPAL data is
shown in Figure \ref{fig:OP_OPAL}.  The agreement is seen to be quite acceptable though one should remember the scale. There are 
differences by factors of up to 2 which actually appear though they are rare.  Similar comparisons were made by
\citet{DelahayePi2005} and \citet{Seaton2007} with data taken from \citet{RicherMiRoetal98} with similar agreement.
See for instance Figure 8 of \citet{Seaton2007} for Si.
However those authors note greater disagreement with  comparisons they make in solar models with \citet{TurcotteRiMietal98}.  The greater differences appear to come mainly from the correction factors discussed above\footnote{There is however one error which we found in Figure 11 of \citealt{TurcotteRiMietal98}:
the curve for C was plotted incorrectly.  It does not correspond to the values used in the calculations.  
This error is responsible for the largest discrepancy found by \citet{DelahayePi2005} and \citet{Seaton2007}.}.  These are always included in the stellar evolution calculations we made and so are included in the figures of  \citet{TurcotteRiMietal98} where \citet{DelahayePi2005} and \citet{Seaton2007} took data for their comparisons.
If, for instance, one looks at Figure 11 of \citet{Seaton2007}, the difference between OP with mte and OPAL is mostly
caused by correction factors shown on Figure \ref{fig:correction} above since 
``mte'' contains only part of the corrections we include.
The correction factors are \emph{not} included in Figure~1 of \citet{RicherMiRoetal98} from which figure \citet{DelahayePi2005} and \citet{Seaton2007} took the data for comparison with that paper.

The OP data has also been used to obtain semianalytic formulas for \gr{} (\citealt{AlecianLe2000,AlecianLe2002}; \citealt{AlecianAr90}; \citealt{LeblancAl2004}).
These require much less computing power than required by integrating over spectra.  They have the further advantage 
of allowing an evaluation for species for which data is not available by using trends in spectroscopic properties. 
These are however less accurate than the detailed evaluation from OP or OPAL and do not allow to take into account the 
 effect of individual concentration variations on Rosseland averaged opacity nor on \gr{}s. 

\section{Examples of the role of  radiative accelerations in Pop I and II stars}
\label{sec:Role}
Stellar evolution calculations including the effect of \gr{}s have now been done for a large number of stars of both Pop I and Pop II.  It is beyond the scope of this brief review to mention all effects of \gr{}s.  However we  briefly describe one effect found in Pop I stars and one in Pop II stars before giving some more details of recent results for HB stars (\S \ref{sec:HBStars}).

The largest structural effect of \gr{}s in Pop I stars is the appearance of an \Fe{} convection zone in all solar metallicity stars more massive than about 1.5 \Msol{} (see \citealt{RichardMiRi2001}).  Iron contributes most to opacity at $\log T \sim 5.3$. Its radiative acceleration  
pushes iron from deeper in the envelope to the point where the \gr(Fe) starts decreasing.  There Fe accumulates during evolution (see Fig. [4] of \citealt{RichardMiRi2001}) approximately where it contributes most to Rosseland opacity.  
This leads to an important increase of  the radiative gradient which, as the Fe abundance increases, becomes larger than the adiabatic gradient and an Fe convection zone appears.  In a 1.5 \Msol{} star this takes a significant fraction of the \MS{} life to occur.  In stars of  1.7 \Msol{} and more, this occurs very early in the \MS{} life.
In stars with $Z = 0.01$, Fe convection zones start appearing at 1.3 \Msol{}.  

The detailed treatment of the interaction of metals with H and He in diffusion processes and of the effect of concentration changes on Rosseland opacity has also shown that an accumulation of metals occurs just outside convective cores and causes semi-convection there (see \S 4 of \citealt{RichardMiRi2001})

The only requirement is for the region with $\log T \sim 5.3$ to be stable enough for diffusion processes to occur there.
There is no adjustable parameter in those calculations.

In Pop II stars, we were surprised to find that \gr{}s and gravitational settling cause abundance anomalies by  factors of 2--10 in a low metallicity cluster such as M92, much larger than those expected in higher metallicity clusters such as M5, M71 or 47 Tuc.  Even in NGC 6397 which is only a factor of 2 more metal rich than M92, the effects are expected to be much smaller though
they have apparently been seen in NGC 6397 \citep{KornGrRietal2006}. In M92, anomalies might have been seen near the turnoff  by \citet{KingStBoetal98} but the noise was large enough to make this uncertain.  This result would need confirmation.

\section{Diffusion in HB stars}
\label{sec:HBStars}
A 0.8 \Msol{} model with $Z = 10^{-4}$ was evolved from the ZAMS through  the core expansion phase on the HB.
\begin{figure}[t!]
\resizebox{\hsize}{!}{\includegraphics[clip=true]{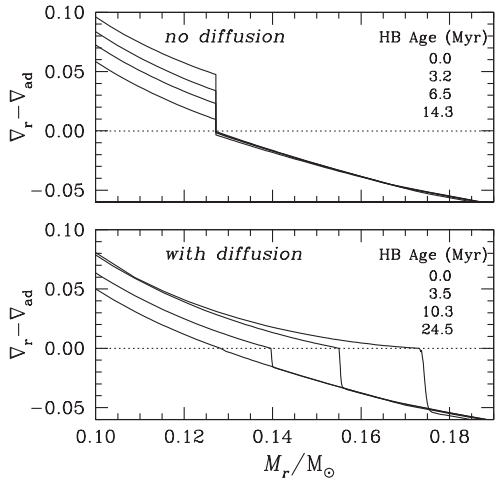}}
\caption{
\footnotesize
Difference between the radiative and adiabatic gradient around the He burning core boundary. 
In both panels, the lowest curve is at zero age HB and the uppermost curve is the final one calculated.
In the calculations with diffusion (lower panel) the convective neutrality condition is automatically satisfied
while it is not satisfied in the absence of diffusion.  Overshooting is not included in these calculations.}
\label{fig:convec_coeur}
\end{figure}
Atomic diffusion was included throughout evolution and no adjustable parameter was involved.  The results were compared to a similar model without diffusion \citep{MichaudRiRi2007}. 

It was found that atomic diffusion had little effect on age determinations using the HB luminosity but that there appeared two interesting effects on the structure of HB stars.  After the discussion of instabilities at the junction of He burning cores and radiative
zones presented by \citet{Paczynski70}, it became accepted that the mixed C cores of HB stars were extended by overshooting or penetration to maintain convective neutrality at the boundary. It is currently not possible to evaluate reasonably accurately the efficiency of overshooting in HB stars. To quote \citet{Sweigart94},``Unfortunately the efficiency of convective overshooting under
such circumstances is entirely unknown. Canonical HB theory assumes that the convective
overshooting is highly efficient...''.  
But as may be seen from Figure \ref{fig:convec_coeur}, it was found that the presence of overshooting during the core expansion phase is made unnecessary by the presence of atomic diffusion.  It was not shown that overshooting plays no role but rather that it is not necessary to invoke such a little understood process. 

Diffusion also causes  an extension of the H burning shell into the He core.  Through the outward diffusion of C from the core and the inward diffusion of H into the core there appears an additional H burning region 
inside the He core which 
 was called diffusion induced H burning (see Fig. [6] of \citealt{MichaudRiRi2007}).  This leads to a slight increase in the ZAHB luminosity.

\subsection{Surface abundance anomalies}
\label{sec:SurfaceAbundanceAnomalies}
Since HB stars are Pop II stars that just left the  giant branch of globular clusters, they are all expected to have the same 
concentration of metals, at least of those heavier than Al \citep{GrattonSnCa2004}.  The concentration of CNO and other relatively light species might show small variations but Fe is not expected to be affected.  \citet{MichaudVaVa83} however suggested that \gr{}s should
lead to overabundances of at least some metals in those stars where settling causes underabundances of \He.  \citet{GlaspeyMiMoetal89}
have confirmed the overabundance of Fe in one star of one cluster but at the limit of detection and this observation required confirmation.
This prediction has now been confirmed in many clusters \citep{BehrCoMcetal99,MoehlerSwLaetal2000,FabbianReGretal2005,PaceRePietal2006} but in particular by \citet{Behr2003} for M15. Overabundances of Fe by factors of 50--100 are seen in all HB stars with $\teff > 11500$\,K while the cooler ones have
the same Fe abundance as cluster's giants. 

\begin{figure}[t!]
\includegraphics[width=\hsize]{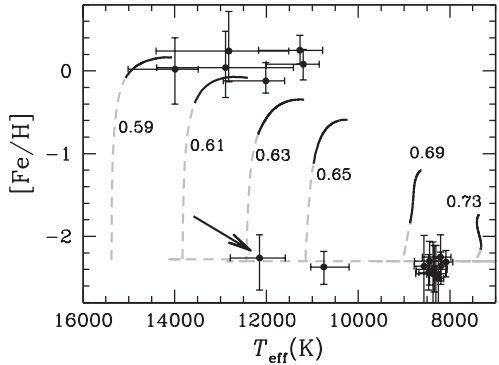}
\caption{
\footnotesize
Concentration of surface [Fe/H] expected in our HB models compared to observations of \citet{Behr2003}
for M15.
The continuous dark lines cover the interval from 5 to 30 Myr after zero age HB for a number of models whose mass is identified on the figure in \Msol.  The star marked with an arrow has $V \sin i \sim 16$ km s$^{-1}$.
All other stars with $\teff > 11000$\,K have $V \sin i <$\,8\,km s$^{-1}$.  Adapted from \citealt{MichaudRiRi2008}.}
\label{fig:Fe}
\end{figure}

In stellar evolution calculations, the surface concentrations depend on the exterior boundary conditions.   In the calculations of \citet{MichaudRiRi2007}, the simplest assumption was made, that of a mixed outer zone without any mass loss.  The mixed mass was adjusted to reproduce approximately the observations of Fe in one of the stars observed by \citet{Behr2003} in M15.  The same model reproduced reasonably well the observations in other high \teff{} stars of that cluster as may be seen in Figure \ref{fig:Fe}.  
Furthermore, as may be seen in \citet{MichaudRiRi2008} the other anomalies are also reasonnably well reproduced.
This is a striking confirmation of the role of \gr{}s in HB stars.

\section{Conclusions}
\label{sec:con}
The availability of large atomic data bases has allowed, as described in \S\,\ref{sec:calculations},  to calculate stellar evolution models for Pop I and II stars up and past the giant branch including all effects of atomic diffusion.
They cause abundance anomalies not only in Ap stars, as originally suggested by \citet{Michaud70}, but also
in HB stars of clusters (\S \ref{sec:SurfaceAbundanceAnomalies}) and possibly in turnoff stars (\S \ref{sec:Role}). They play an essential role in driving pulsations in sdB stars \citep{FontaineBrChetal2003}, the field analogue of HB stars. It is furthermore not only the surface region which is affected but 50\% of the stellar radius and 10$^{-3}$ 
of its mass \citep{RichardMiRi2002,MichaudRiRi2007}.

\citet{Eddington26} was right however in suggesting that competing processes also have a role to play.  For instance, in M15, rotation plays a role probably through meridional circulation. The arrow in Figure \ref{fig:Fe} shows the only star with $\teff> 11500$\,K which has no abundance anomaly.  It is also the only relatively rapidly rotating star.  \citet{QuievyChMietal2007} have shown that meridional circulation explained that the stars with $\teff<11500$\,K have a normal abundance and not the 5$\times$ overabundance that Figure \ref{fig:Fe} would suggest.  While atomic diffusion driven by \gr{}s plays the main role in creating abundance anomalies on the HB, it has to compete with the effects of rotation just as in HgMn or AmFm stars \citep{CharbonneauMi91}.

\begin{acknowledgements}
This research was partially supported at  the Universit\'e de Montr\'eal 
by NSERC. We thank the R\'eseau qu\'eb\'ecois de calcul de haute
performance (RQCHP)
for providing us with the computational resources required for this
work. 
\end{acknowledgements}

\bibliographystyle{aa}



\end{document}